\documentclass[10pt, twocolumn,conference]{IEEEtran}
\IEEEoverridecommandlockouts

\usepackage{cite}
\usepackage{amsmath,amssymb,amsfonts}
\usepackage{algorithmic}
\usepackage{algorithm}
\usepackage{graphicx}
\graphicspath{{./epsfiles/}}
\usepackage{textcomp}
\usepackage{xcolor}
\usepackage{booktabs}
\usepackage{xcolor}
\usepackage{colortbl}

\usepackage{graphicx}
\usepackage{subfigure}

\usepackage{adjustbox}

\usepackage[a4paper, left=1.75cm, right=1.75cm, bottom=2.54cm, top=0.79in]{geometry} 
\begin{document}

\title{Digital Twin-Enhanced Deep Reinforcement Learning for Resource Management \\ in Networks Slicing
}

\author{Zhengming Zhang,~\IEEEmembership{Student Member,~IEEE},   Yongming Huang, ~\IEEEmembership{Senior Member,~IEEE}, Cheng Zhang,\\ ~\IEEEmembership{Member,~IEEE}, Qingbi Zheng, Luxi Yang, ~\IEEEmembership{Senior Member,~IEEE}, Xiaohu You, ~\IEEEmembership{Fellow,~IEEE}}

\maketitle

\begin{abstract}
Network slicing-based communication systems can dynamically and efficiently allocate resources for diversified services. 
However, due to the limitation of the network interface on channel access and the complexity of the resource allocation, it is challenging to achieve an acceptable solution in the practical system without precise prior knowledge of the dynamics probability model of the service requests. 
Existing work attempts to solve this problem using deep reinforcement learning (DRL), however, such methods usually require a lot of interaction with the real environment in order to achieve good results.
In this paper, a framework consisting of a digital twin and reinforcement learning agents is present to handle the issue.
Specifically, we propose to use the historical data and the neural networks to build a digital twin model to simulate the state variation law of the real environment. Then, we use the data generated by the network slicing environment to calibrate the digital twin so that it is in sync with the real environment. Finally, DRL for slice optimization optimizes its
own performance in this virtual pre-verification environment. We conducted an exhaustive verification of the proposed digital twin framework to confirm its scalability. Specifically, we propose to use loss landscapes to visualize the generalization of DRL solutions.  We explore a distillation-based optimization scheme for lightweight slicing strategies. In addition, we also extend the framework to offline reinforcement learning, where solutions can be used to obtain intelligent decisions based solely on historical data.
Numerical simulation experiments show that the proposed digital twin can significantly improve the performance of the slice optimization strategy. 
\end{abstract}

\begin{IEEEkeywords}
Network slice; digital twin; deep reinforcement learning; resource management.
\end{IEEEkeywords}

\section{Introduction}
\renewcommand{\thefootnote}{}
\footnotetext{
Z. Zhang, Y. Huang, C. Zhang, L. Yang, and X. You are with the School of Information Science and Engineering, and the National Mobile Communications Research Laboratory, Southeast University, Nanjing 210096, China, and also with the Pervasive Communications Center, Purple Mountain Laboratories, Nanjing 211111, China (e-mail: zmzhang@seu.edu.cn; huangym@seu.edu.cn; lxyang@seu.edu.cn, zhangcheng$\_$seu@seu.edu.cn, xhyu@seu.edu.cn).

Q. Zheng is with the Future Research Lab, China Mobile Research Institute, Beijing 100053, China (e-mail: zhengqingbi@chinamobile.com).
}

With the explosive growth in the number of mobile users and applications, novel network architecture and emerging technologies are expected to offer support for multifarious network services with diverse performance requirements \cite{katsalis2017network}. 
The key technologies of 5G wireless systems spawn three generic application scenarios: enhanced mobile broadband (eMBB), massive machine-type communications (mMTC), and ultra-reliable and low-latency communications (URLLC) \cite{series2017minimum}. Moreover, 6G will further expand connectivity and service coverage to achieve extreme connectivity performance with Tbps-scale data rate, Kbps/Hz scale spectral efficiency, and $\mu$s-scale latency \cite{10183795}. It is foreseeable that the upcoming 6G will provide the ultimate quality of service (QoS) for a variety of services (such as virtual reality and vehicle-to-infrastructure communications).
However, traditional mobile networks are mostly designed to serve mobile broadband users and contain only a few tunable parameters, such as the prioritization of dedicated services and QoS \cite{hua2019gan}. Due to the different service requirements for network design and development, it is difficult for mobile operators to extend their networks into these emerging vertical services.
Hence, network slicing \cite{7926923,8004168} is proposed as a core feature of advanced networks that allows the deployment of multiple logical networks over a common physical infrastructure to address the above challenging problems. In network slicing, each logical network provides a specific type of service. These logical networks referred to as network slices, can be managed and exploited by third-party entities or tenants that slice physical and computational resources of the network to meet the diverse needs of a range of users \cite{zhou2016network,li2017network}.

However, in the radio access network (RAN), in order to fully exploit the power of network slicing to provide better performance and cost-effective services for communication networks, RAN slicing involves more challenging technical issues for existing resource management, because for RAN, spectrum is a scarce resource, ensuring spectral efficiency (SE) is critical \cite{zhang2017network}, and the level of assurance of service per user usually imposes strict requirements, in addition, the actual demand for each slice largely depends on the request pattern of mobile users \cite{hua2019gan}.
Starting with the attention that network slicing received in wireless communications, many efforts began to try to solve the above problems. In \cite{8370043}, the RAN slicing problem is formulated as a bi-convex problem. Two algorithms are designed to address the bi-convex problem. However, in their solutions, the explicit relationship between required resources and service level agreements (SLAs) on network slices is not considered. In \cite{8761841}, the resource management of RAN is studied from two aspects: user access control and wireless bandwidth allocation. However, its solution relies on the assumption that different users have the same fixed demand rate, an assumption that may not be met in practice. 
In \cite{8070468} the queuing model is used to minimize transmission power, and bandwidth to achieve a high-quality URLLC slice algorithm. In \cite{8638932}, the objective function of the considered network slicing problem is transformed into a mixed integer nonlinear programming, which is solved by continuous convex approximation and semidefinite relaxation method.

However, current researches \cite{9259378,8962338,9852968} point that traditional methods require precise mathematical models and known parameters, which are often difficult to achieve in practice. Usually, dynamic programming (DP)~\cite{Ref20,Ref21} strategies are used to tackle this type of decision-making problem. However, DP requires the knowledge of the precise state transition function (e.g., users' demands queue model), and due to the curse of dimensionality, this type of algorithm can only be used for problems with limited action space. 
Intelligent algorithms (e.g., learning-based solutions) do not require many assumptions and do not require prior information about the system are expected to solve the decision-making problem of network slicing.
For example, deep reinforcement learning (DRL) which utilizes the powerful function approximation ability to solve the above problem has attracted a lot of research. 
DRL is a field of machine learning that sets an agent to adopt a policy in an environment (e.g., a RAN with network slicing deployed), while the policy is a sequence of actions that the agent performs according to a learned model in order to maximize the cumulative reward during its interactions with the environment.

DRL learns the probability of state transitions during interaction with the environment and the function-fitting ability of the neural network could be used to handle large-scale action space problems. DRL has triggered tremendous research attention to solving resource allocation issues, e.g., power control \cite{9810823}, beam tracking \cite{9410605}, and offloading \cite{8620546}.
As for network slicing, DRL is used in \cite{8540003} to solve some resource management problems in typical network slicing scenarios, including radio resource slicing and priority-based core network slicing, and simulation verifies the advantages of DRL in some competitive schemes. 
In \cite{10280728}, a hierarchical intelligent resource management method based on a deep neural network and a multiarm bandit (MAB) under a hierarchical intelligent controlling framework for RAN is proposed.
In \cite{hua2019gan}, a generative adversarial network-driven deep distribution Q network is proposed to learn the action-value distribution by minimizing the difference between the estimated action-value distribution and the target action-value distribution, thereby reducing the impact of annoying randomness and noise embedded in the receiving SLA satisfaction ratio (SSR) and spectral efficiency (SE). In \cite{9616406}, the allocation of power and radio resources are considered, and deep learning is used for decision-making on resource allocation on a large time scale while DRL is used for decision-making on resource allocation on a small time-scale. DRL is also used in \cite{9791425} and \cite{9616406} to satisfy QoS requirements while conformer long-term slicing policies must be, no matter how unbalanced they are. Although the above work has shown us the effectiveness of DRL for network slicing problems, there are still challenges in applying DRL-based solutions to real-world systems. A key problem is that DRL requires a lot of interaction with the environment to learn and enhance its performance through trial and error, and such algorithms often require a large number of samples, which involves extensive exploration of strategies, including inefficient ones. However, the cost of interacting with the actual environment is high, and the overhead of collecting large amounts of data is enormous. A virtual pre-validation environment tailored to network slicing problems is urgently needed to optimize DRL performance.

The motivation of our work is to explore digital twin-based schemes to build a virtual pre-validation environment (a twin of the real environment) and improve the learning efficiency in the process of reinforcement learning interacting with the real environment. 
Digital twin technology \cite{9899718,DT} provides a new idea for the construction of a new wireless network architecture. Most of the current network slicing solutions need to rely on a priori assumptions about the network environment or a large number of interactions with the wireless communication environment. However, these assumptions often cannot fully adapt to the actual network environment. In the process of a large number of interactions between the algorithm and the environment, the cost of trial and error is high, and there is a risk of causing huge fluctuations in network performance. This paper argues that the digital twin of the wireless communication system can be constructed, making full use of real network environment data, integrating multi-physical quantities and multi-scale simulation processes, and completing the mapping in the virtual space, thus reflecting the life cycle process of the physical network. The digital twin adopts a highly flexible and repeatable development method to cost-effectively design and verify wireless network slicing systems, thereby reducing development difficulty and testing risk, and improving research and operational efficiency.

In order to leverage the capabilities of digital twins mentioned above, it is natural that we need to provide potential solutions to the following three problems:

Q1: How to build a digital twin that maps the real network slicing environment to the digital domain?

Q2: How to synchronize the digital twin with the real slicing environment so that the digital twin matches the real environment? 

Q3: How to use the digital twin to enhance the performance of network slicing?

In this paper, we provide solutions to the above three problems, the details of which are presented in Section III.
Specifically, we propose a virtual-real co-existence technology between the digital twin and the actual slicing environment and establish a corresponding closed-loop control system. More specifically, we consider how to allocate RAN resources in a software-defined networking-based system, maximizing the resources available to infrastructure providers while securing SLAs for hosting tenants. In order to achieve the above resource management goals, we propose to build a digital twin for network slicing using historical data. Then, the DRL-based network slicing policy is optimized with the digital twin to improve efficiency and performance.

In this context, our main contributions are the following:

1) We propose a method to construct digital twins from the historical data of network slicing systems to answer Q1. This digital twin is trained in a supervised manner, learning from the dynamics of the real network slicing system through deep neural networks. The trained digital twin is a generative model that generates system behavior that matches the actual network slicing environment. 

2)  A digital twin-enhanced DRL framework for network slicing optimization is proposed to answer Q2 and Q3. The framework consists of two closed loops, which are the outer loop and the inner loop. The outer loop provides fresh data to the digital twin to calibrate it so that it is in sync with the real environment. In the inner loop, the digital twin is used as a virtual pre-validation environment for the DRL agent's optimization.

3) Furthermore, we extend the proposed digital twin enhanced DRL network slicing method to further demonstrate the advantages of digital twins.
Specifically, we combine the digital twin-optimized network slicing module with knowledge distillation to compress the intelligent network slicing model into small models with low computing complexity. Besides, we propose to use the digital twin module to enhance the performance of offline reinforcement learning-based slicing algorithms which learn good control policies only from old data that can be gleaned from previous control policies.

4)  We conduct experiments to find resource allocation strategies under the uncertainty of slicing service requirements in a software-defined network (SDN)-enabled slice networks. We compare and analyze quantities of the DRL baselines and proposed digital twin-enhanced DRL solutions, namely loss landscape \cite{li2018visualizing}. We find that digital twin-enhanced reinforcement learning has a flatter loss landscape than the baseline (e.g., DDQN solution), which indicates that the digital twin can improve the robustness and generalization ability.

Besides, the simulation results show that adding the proposed digital twin module to the network slicing system based on reinforcement learning can improve the performance of the model.
In general, the digital twin model proposed by this work has a certain versatility, which can provide new ideas for efficient and secure smart wireless network slicing, which is worth further exploration.

The remainder of this paper is organized as follows. Section II introduces the system model and formulates the resource management problem in the network slicing problem. Section III presents the proposed digital twin-enhanced deep reinforcement learning approach. Section IV evaluates the performance of the proposed approach through simulation results. Section V concludes this paper.

\section{System Model, Resource Management in Network Slicing and the Deep Reinforcement Learning-Based Solution}
In this section, first, we introduce the considered resource management problem in network slicing and then we present the DRL framework for the considered problem.

\subsection{System Model}

\begin{figure}
    \centering
    \includegraphics[width=0.998\linewidth]{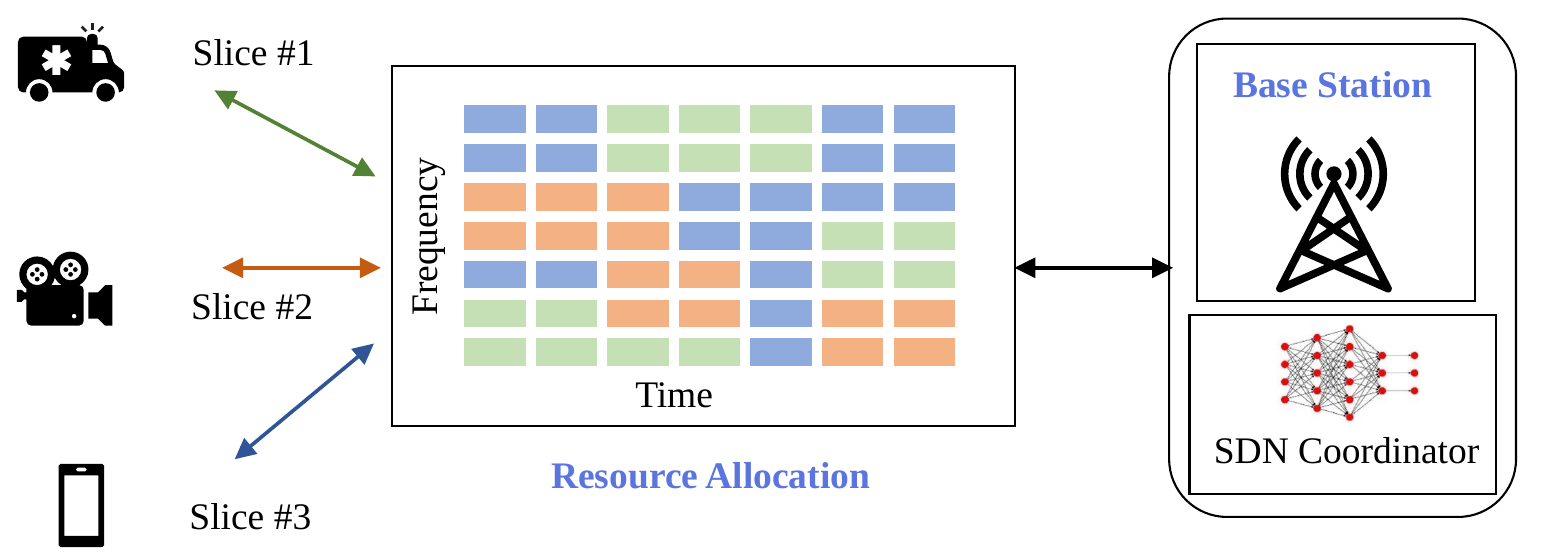}
    \caption{Diagram of the system.}
    \label{Systemmodel}
\end{figure}

We consider the SDN-based system model for the dynamic
allocation of wireless bandwidth in the RAN scenario with
downlink transmissions. The system model is shown in Fig.~\ref{Systemmodel}. In the framework of the considered hierarchical network slices, we consider a RAN scenario with a single base station (BS) where $N$ network slices share bandwidth. A base station, covering users on different RAN slices, transmits downlink signals in the cells it covers. Each RAN slice is granted a predefined subset of bandwidth resources in each time-frequency frame. During one observation period, the SDN control system monitors the performance and updates the bandwidth resource allocation for the next observation period based on the observed variables of the slice and the quality indicators of the slice service.
For the $n$-th network slice where $n\in {\cal N}$, it provides a single service for a set of users ${\cal{K}}_n$. In the system under consideration, slicing decisions are made periodically. In a time slot, the number of requests received by the $n$-th network slice is denoted as $d_n$ which partially determines the wireless bandwidth $w_n$ allocated by BS to this $n$-th network slice. The same as \cite{hua2019gan}, $d_n$ depends on both the number of demands and the bandwidth-allocation solution in the previous timeslot.

\subsection{Resource Management Problem}
The goal of resource management in the considered slicing scenario is to find a suitable bandwidth allocation scheme that maximizes the utility of the system. The utility is a synthesis of the users' spectrum efficiency and SLA satisfaction ratio, i.e. SSR, which is often considered to be the weighted sum of the two.

Let $\chi_{n,k}$ be the signal noise ratio (SNR) of user $k\in {\cal{K}}_n$. It is calculated by the following equation:
\begin{equation}
{\chi _{n,k}} = \frac{{{g_{n,k}}{P_{n,k}}}}{{{N_0}{w_n}}},    
\end{equation}
where ${g_{n,k}}$ is the channel gain from the BS to the user $k$, ${P_{n,k}}$ is the
transmission power and $N_0$ is the noise power. Assume the total bandwidth is $W$ and the user rate is $r_{n,k}$ of the user $k$, the spectrum efficiency $s$ is defined as:
\begin{equation}
\left\{ \begin{array}{l}
s = \sum\limits_{n \in {\cal N}} {\sum\limits_{k \in {{\cal K}_n}} {\frac{{{r_{n,k}}}}{W}} }, \\
{r_{n,k}} = {w_n}\log (1 + {\chi _{n,k}}).
\end{array} \right.    
\end{equation}

The SSR of $n$-th network slice is obtained by dividing the number of successfully transmitted packets by the total number of arrived packets.
That is for $n$-th network slice the SSR $u_n$ is:
\begin{equation}
{u_n} = \frac{{\sum\limits_{k \in {{\cal K}_n}} {\sum\limits_{{q_k} \in {Q_n}} {{x_{n,k}}} } }}{{\sum\limits_{k \in {{\cal K}_n}} {\left| {{Q_{{n}}}} \right|} }},   
\end{equation}
where ${Q_n}$ is the set of packets sent from the BS to user $k\in {\cal{K}}_n$, ${\left| {{Q_{{n}}}} \right|}$ is the cardinality of the set ${Q_n}$ and ${{x_{n,k}}}\in \{0,1\}$ where ${{x_{n,k}}}=1$
indicates that the packet ${q_k} \in {Q_n}$ is successfully received by the user, i.e., the downlink data rate and the latency are simultaneously satisfied the predetermined rate and latency requirement according to the SLA for service type $n$. If the rate or latency requirements are not met, ${{x_{n,k}}}$ is set to 0.
The bandwidth allocation problem in the long-term RAN network slice performance optimization is formulated as:
\begin{equation}
\begin{array}{l}
\mathop {\max }\limits_{w_n^t} \frac{1}{T}\sum\nolimits_{t \in T} \sum\nolimits_{n \in {\cal N}} {\left( {\alpha s(w_n^t) + {\beta_n {u_n}(w_n^t)} } \right)}, \\
s.t.\sum\limits_{n \in {\cal N}} {w_n^t}  = W,\\
\quad \ \sum\limits_{n \in {{\cal K}_n}} {\left| {{Q^t_n}} \right| = {d^t_n}}, \\
\quad \   {x_{n,k}} \in \{ 0,1\},
\end{array}    
\label{SliceProblem}
\end{equation}
where $\alpha$ and $\beta_n$ are the coefficients that adjust the importance of spectrum efficiency $s$ and SSR $u_n$. These coefficients can be set to any positive number, e.g., \cite{hua2019gan} sets $\alpha=0.01$ and $\beta_n=1, \forall n \in {\cal N}$ in the simulation experiment. In the timeslot sequence, since the maximum transmission capacity of the RAN belonging to a service is entangled with the service's ability to supply, the change in traffic demand $d^t_n$ is affected by user demand and bandwidth allocation decisions. The traffic demand $d^t_n$ varies without knowing a prior transition probability, making the above problem (\ref{SliceProblem}) difficult to solve.

Specifically, the challenges of solving the resource management problem (\ref{SliceProblem}) are mainly reflected in the following aspects:

1) It is difficult to obtain a solution without accurately obtaining the probability distribution model behind the environment, such as the distribution model of user requests and network traffic.

2) The feasible space where the decision variable $w_n$ of resource management is located is huge, and standard brute force algorithms are often impractical due to the curse of the action space dimension.

In order to solve the resource management problem in the dynamic network slicing environment, the deep reinforcement learning solution can be considered as a promising approach.

\subsection{Deep Reinforcement Learning-Based Network Slicing Optimization Module}

Here, we introduce network-slicing optimization modules based on deep reinforcement learning.
Specifically, for our resource management problem, the definitions of each tuple $\left\langle S_{,} A, P, R, T\right\rangle$ are

(1) State space (${\cal S}$): In stage $t$, the state, $s_{t} \in {\cal S}$ is the number of packets arriving per slice in a specific time window.

(2) Action space (\emph{A}): \emph{A} is a limited set of all alternative actions. The action~$a_{t} \in A$ is the bandwidth allocation for each slice, e.g., $a_{t}=\{w^t_n\}_{n\in {\cal N}}$.

(3) State transition probability (\emph{P}):~$\emph{Pr}\left[s_{t+1} \in{ \cal S} \mid s_{t,} a_{t}\right]$ is the probability from~$s_t$ to~$s_{t+1}$ after the selection of an action~$a_{t} \in A$. Since there is a nonlinear relationship between the available bandwidth resources, the user demands, and the SSR, the state transition probability can not be given by an accurate expression.

(4) Reward function (\emph{R}): The reward~$r_t$ is defined as a function of utility, where the utility is defined as:
\begin{equation}
{U^t} = \sum\nolimits_n {\alpha s(w_n^t) + \sum\nolimits_{n} {\beta_n{u_n}(w_n^t)} },
\label{utility}
\end{equation}
and the reward $r_t$ is a reshaping function the same as \cite{hua2019gan}:
\begin{equation}
\left\{ \begin{array}{l}
{r_t} = 0,{U^t} < {\gamma ^t},\\
{r_t} = 1,{\gamma^t} \le {U^t} < {\gamma _{\max }},\\
{r_t} = 1 + ({U^t} - {\gamma _{\max }})/2,otherwise,
\end{array} \right.
\label{reward_equ}
\end{equation}
where ${\gamma ^t}$ is a threshold defined as 
${\gamma ^t} = \min \{ {\gamma _{\min }} + \sigma t,{\gamma _{\max }}\}$ and $\sigma$ is the linear growth coefficient of the threshold.

(5) Time Steps (\emph{T}): \emph{T} is the set of slice policy update time period. 
We define the transition from~$S_t$ to~$S_{t+1}$ as one step within 400 scheduling slots and each scheduling slot lasted 0.5ms. 

Thus, our goal is to obtain more reward $r_t$ in all time steps, i.e.,~${\rm{maxmize}}\sum\nolimits_t {{\lambda^t r_t}}$ where $\lambda^t$ is the discount factor. 
Although DRL does not need to know the specific mathematical model of the network slicing problem, it can optimize the slicing strategy simply by trial and error. However, it requires a lot of interaction with the environment when it is actually used, and its low learning efficiency often limits its practical value. Therefore, this paper proposes a DRL training framework enhanced by a digital twin to improve performance.

\section {Digital Twin-Enhanced Deep Reinforcement Learning Solution}

Model-free reinforcement learning using deep learning architecture achieves good performance on a variety of complex tasks. 
However, such algorithms need to be learned through trial and error. The limitation of such approaches is that they often require a lot of interaction with the environment for training. When these interactions are costly or even dangerous, the usefulness of such algorithms is greatly reduced. 

Here, we introduce a digital twin-enhanced DRL framework to solve the above problems. The core of this framework consists of digital twins and slice-optimized agents. We first introduce this framework, and then introduce the construction of digital twins (Q1's solution), the synchronization of digital twins with real environments (Q2's solution), and the optimization of slicing strategies based on digital twins (Q3's solution).

\subsection{Digital Twin-Enhanced DRL Framework} 

\begin{figure*}
    \centering
    \includegraphics[width=0.85\linewidth]{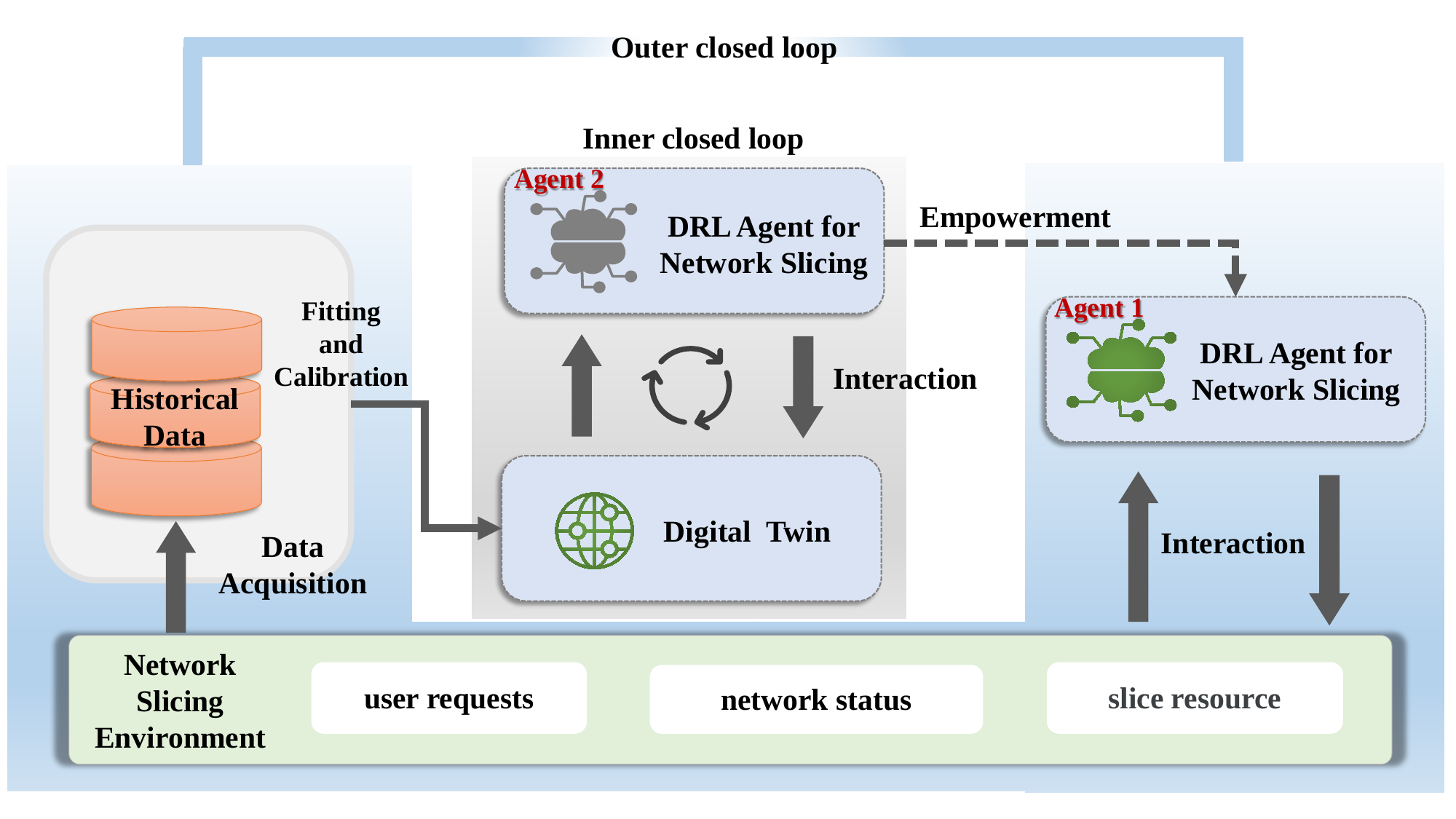}
    \caption{Diagram of the digital twin-enhanced deep reinforcement learning framework.}
    \label{framework}
\end{figure*}

\begin{algorithm}
\caption{Digital Twin-Enhanced Deep Reinforcement Learning Solution}
\label{alg2}
\begin{algorithmic}[1]
 \STATE  \textbf{\underline{Initialize:}} Initialize the DRL model of Agent 1 with random parameters $\theta ^{1,1}_{1}$, set the initial parameters $\theta ^{1,1}_{2}$ of the DRL model of Agent 2 as $\theta ^{1,1}_{2}=\theta ^{1,1}_{1}$.
 Initialize the LSTM model $P_\theta$ and the neural network model $R_\theta$ with random parameters.
\STATE \textbf{For $I=1,\cdots,\Gamma$:} 
\STATE \quad \textbf{\underline{Data Collection:}} 
Use the data acquisition module with policy $\pi_1$ to collect some data $(s_t,a_t,r_t,s_{t+1})$ and obtain a training dataset ${\cal D}=\{(s_i,a_i,r_i,s_{i+1})\}_{i=1}^{M}$.
\STATE \quad \textbf{\underline{Train the Digital Twin :}} Train the LSTM model $P_\theta$ and the neural network model  $R_\theta$ using the following loss function with the dataset ${\cal D}$:
\begin{equation}
\left\{ \begin{array}{l}
{L_p}(\theta ) = \frac{1}{{M}}\sum\nolimits_{\left( {s,a} \right) \in D} {{{\left\| {{P_{\theta_p} }({s_t},{a_t}) - {s_{t + 1}}} \right\|}^2}}, \\
{L_R}(\theta ) = \frac{1}{{M}}\sum\nolimits_{\left( {s,a,r} \right) \in D} {{{\left\| {{R_{\theta_R} }({s_t},{a_t}) - {r_t}} \right\|}^2}}.
\end{array} \right.
\label{loss_dt}
\end{equation}
\STATE \quad  \textbf{\underline{Train the DRL Agent 2:}} 
\STATE \quad \textbf{For $k=1,\cdots,\Phi$:} \\
\STATE \quad \quad Observe the current state~$s_k$ of step $k$, i.e., the number of packets arriving per slice.
\STATE \quad \quad  According to the policy $\pi_2$ gets the bandwidth allocations $a_k=\{w^k_n\}_{n\in{\cal N}}$ of the current state, i.e.,  select bandwidth allocations~$a_k \! =\! \mathop {\arg \max }\limits_a Q \!(s_k ,a;\theta ^{I,k}_{2}  )$.  \\
\STATE \quad \quad Provide current state $s_k$ and action $a_k$ to ${P_{\theta_p} }$ and ${R_{\theta_R} }$, and observe the predicted new state~$s_{k+1}$ and the predicted reward~$r_k$.
\STATE \quad \quad Store transition~$\left( {s_k ,a_k ,r_k ,s_{k + 1} } \right)$ in a buffer~$\Sigma_2$. Then, sample a random minibatch of \({ N}\) samples from~$\Sigma_2$ and train the Agent model to update $\theta _2^{I,k}$ with a DRL algorithm $\phi$.
\STATE \quad \textbf{End For}
\STATE \quad  \textbf{\underline{Agent 2 Empowers Agent 1:}} 
Set:
\begin{equation}
\theta _1^{I,1} = \zeta \theta _1^{I,T} + (1 - \zeta )\theta _2^{I,\Phi},  
\label{empower_equation}
\end{equation}
where $\zeta$ is a hyperparameter characterizing the strength of empowerment, and we set when $I=1$, $\theta _1^{I,T}=\theta _1^{I,1}$.
\STATE \quad  \textbf{\underline{Train the DRL Agent 1:}} 
\STATE \quad \textbf{For $t=1,\cdots,T$:} \\
\STATE \quad \quad Observe the current state~$s_t$ of step $t$.
\STATE  \quad \quad With probability~$\xi$ select a random bandwidth allocation action $a_t$. Otherwise select~$a_t \! =\! \mathop {\arg \max }\limits_a Q \!(s_t ,a;\theta ^{I,t}_{1}  )$.  \\
\STATE \quad \quad Execute action $a_t$ in the network slicing environment and observe reward $r_t$ and the next state $s_{t+1}$.
\STATE \quad \quad Store transition~$\left( {s_t ,a_t ,r_t ,s_{t + 1} } \right)$ in a buffer~$\Sigma_1$. Then, sample a random minibatch of \({ N}\) samples from~$\Sigma_1$ and train the Agent model with the DRL algorithm $\phi$  to update $\theta _1^{I,t}$.
\STATE \quad \textbf{End For}
\STATE \quad  Copy the parameters of Agent 1 to Agent 2.
\STATE \textbf{End For}
\end{algorithmic}
\end{algorithm}

We propose an architecture as shown in Fig. \ref{framework} which uses a digital twin to address the intelligent network slicing problem.
The architecture contains two closed loops: an inner closed loop and an outer closed loop. The intuition behind the two closed loops is that the digital twin, as a virtual entity in the digital domain of the real environment, can form a closed loop with the DRL slicing optimization module, and the real environment can also form a closed loop with the DRL slicing optimization module.

The compositions of these two closed loops are described as follows:

1) The outer closed loop mainly includes the network slicing environment, a data acquisition module, and a DRL agent for slice performance optimization (we refer to this DRL agent as Agent 1).

2) The inner closed loop includes the digital twin module and another DRL agent (we refer to this DRL agent as Agent 2) that does not interact with the environment in the outer closed loop. 

The two closed loops are connected through a data acquisition module. Specifically, the data acquisition module provides data support for the fitting and calibration of the digital twin and synchronizes the digital twin with the real environment. Besides, the concrete operation of the framework is shown in Algorithm 1.

\subsection{Data Acquisition Module}
Building a digital twin requires collecting a certain amount of data. The data acquisition phase (line 3 in Algorithm 1) of the intelligent RAN slicing system needs to be completed:

1) Collect slice resource configuration data from the base station, such as the number of bandwidth resources; 

2) Collect network status parameters from the base station, such as the number of packets arriving in each slice within a specific time range;

3) Collect slicing performance requirements parameters from the base station, such as user spectrum efficiency, delay, and the number of packets transmitted. As for the slice performance statistics, the method is the time window statistical method, that is, the time window is used as the time dividing line to count the slicing performance parameters.

\subsection{The Construction of Digital Twins}

In the inner closed loop of the proposed framework, the intelligent slicing algorithm learns the slicing strategy by interacting with the verification environment model instead of the real environment to minimize the loss of the actual system brought by exploration. 

Here, we give the solution for Q1: 
The digital twin, as a virtual, pre-validated environment model, is trained in a supervised manner, learning from the dynamics of the real system through deep neural networks.

\begin{figure}[t]	\centerline{\includegraphics[width=0.5\textwidth]{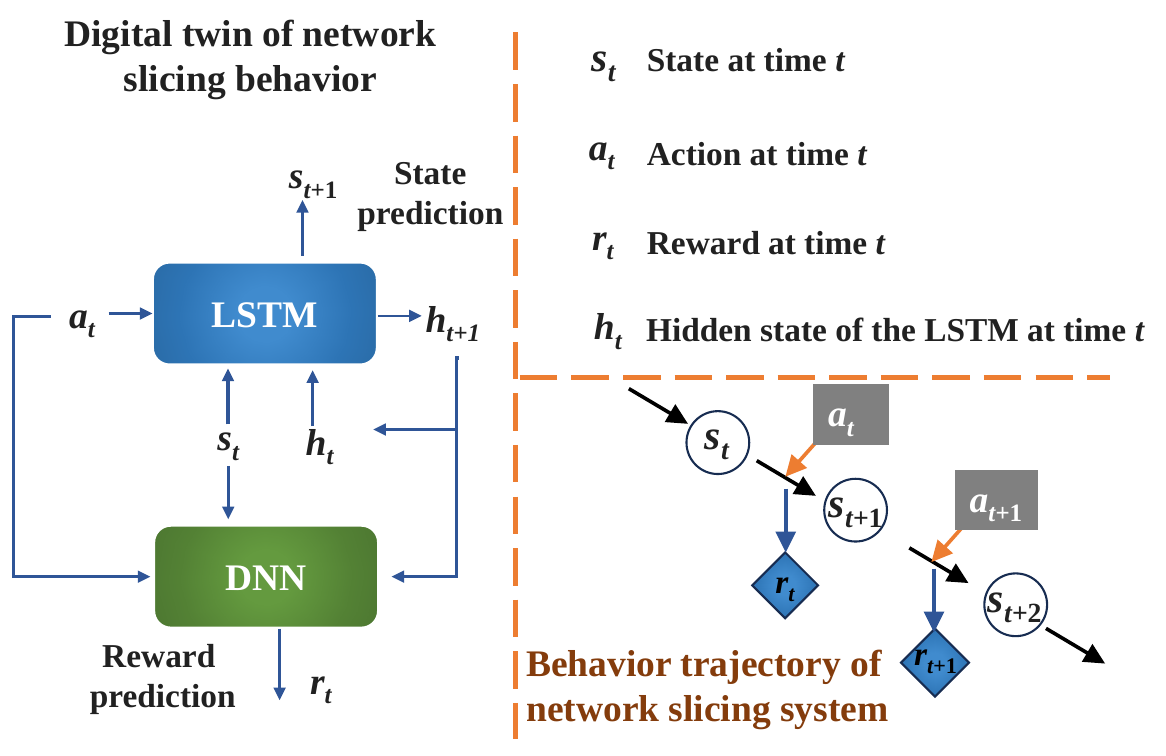}}
	\caption{Diagram of a digital twin that describes the behavior of a network slicing system.}
	\label{behavior_digital_twin}
\end{figure}

Specifically, we design a pre-validation module as the digital twin to learn the environment model. The pre-validation module contains the following two models: one is the state transition prediction model, which enters the current state $s_t$ and action $a_t$ to predict the next state $s_{t+1}$. The other is a reward prediction model, which inputs the current state $s_t$ and bandwidth allocation action $a_t$ to predict the reward $r_t$. That is, the model can be described as the following two formulas:
\begin{equation}
\left\{ \begin{array}{l}
{s_{t + 1}} \sim {P_{\theta_p} }({s_{t + 1}}|{s_t},{a_t}),\\
{r_{t }} \sim {R_{\theta_R} }({r_{t }}|{s_t},{a_t}).
\end{array} \right.    
\end{equation}
If the above environment model can accurately describe the transformation model of the real environment, then we can realize the prediction of states and rewards based on the model, and when a new state and action arrives, we can directly predict the new state and action reward based on the model, without interacting with the environment. To this end, we design to use a long short-term memory \cite{hochreiter1997long} (LSTM) network  $P_{\theta_p}$ to complete the modeling of the state transition model and a deep neural network (DNN) $R_{\theta_R}$ to complete the fitting of the reward model. The digital twin can be seen as describing the behavior of the network slicing system, as shown in Fig. \ref{behavior_digital_twin}. The behavior of the slicing system in the figure is described by the digital twin as a time series. The network state is transferred to a new state under the bandwidth allocation $\{w^t_n\}_{n\in {\cal N}}$ of the wireless services, and the system obtains the corresponding utility in the process. 

For the synchronization of the digital twin and real environment (Q2's solution), our solution includes the steps of lines 3 and 4 in Algorithm 1. 
Specifically, the steps of performing data collection using $\pi_1$ and model training using (\ref{loss_dt}) are intended to fit and calibrate the digital twin so that the digital twin adapts to the real environment.

One can see that the LSTM-based digital twin is a generative model that generates a sequence of states in an autoregressive manner (specifying that the output state depends on its own previous value and a random term). In our digital twin framework, we are not limited to using LSTM to generate state sequences. Other generative sequence models such as transformers \cite{vaswani2017attention} can also be used as state generators.
While considering that the digital twin for network slicing needs to be able to efficiently interact with the network slicing optimization module, in order to achieve a trade-off between efficiency and performance, we choose to use an LSTM that is smaller than large generative models such as transformers and GPTs \cite{radford2018improving}.

\subsection{DRL Agents}
The goal of a DRL agent is to find a policy/solution $\pi$ that maps states to bandwidth allocations $\{w^t_n\}_{n\in {\cal N}}$, i.e. $\pi(s_t)=a_t$, to maximize an action-value function:
\begin{equation}
{Q_\pi }(s,a) = E\left[ {\sum\nolimits_t {{\lambda ^t}{r_t}} |{s_0} = s,{a_0} = a,\pi } \right].    
\end{equation}
The optimal action-value function ${Q^*}(s,a) = {\max _\pi }E\left[ {\sum\nolimits_t {{\lambda ^t}{r_t}} |{s_0} = s,{a_0} = a,\pi } \right]$ is the maximum expected shaping utility (\ref{reward_equ}) achievable by following the policy $\pi^{*}$. 
Here, we mention three solutions for agents to obtain bandwidth allocations to the slice problem. There are deep Q-network (DQN)~\cite{DQN}, Deep double Q-network approach (DDQN) \cite{DDQN} and Dueling GAN-DDQN \cite{hua2019gan}. Agents trained by these three solutions can serve as DRL modules in our digital twin-enhanced intelligent network slicing framework (Fig. \ref{framework}).

\subsection{Optimization of the Network Slicing Policy with the Digital Twin}

Now, we introduce the solution (Algorithm 1) we provide for the optimization of the network slicing strategy with the digital twin (Q3's solution) in detail below.

For the optimization of the slicing policy, our solution includes the interactive training steps of Agent 2 and the digital twin (lines 5-11 in Algorithm 1), the steps of Agent 2 empowering Agent 1 (line 12 in Algorithm 1), and the steps of Agent 1 accessing the real environment to perform fine-tuning  (lines 13-19 in Algorithm 1). The purpose of the step of Agent 2 empowering Agent 1, i.e., (\ref{empower_equation}), is to synchronize the DRL agent that fully interacts with the pre-verification environment to the agent that interacts with the actual environment. The purpose of Agent 1 continuing to interact with the actual environment for a period of time is to fine-tune its parameters to better adapt to the network slicing environment.

We repeat the above process in Algorithm 1 until Agent 1 converges. Besides, the DRL algorithm $\phi$ (lines 10 and 18 of Algorithm 1) can be selected from DQN, DDQN, and Dueling GAN-DDQN, and different digital twin-enhanced DRL algorithms can be obtained accordingly. For example, one can choose DQN as $\phi$ and get a digital twin-enhanced DQN algorithm. The simulation part of this paper will prove that the performance of network slicing can be improved after adding the proposed digital twin module to any algorithm in DQN, DDQN, and Dueling GAN-DDQN and obtaining the corresponding digital twin-enhanced algorithm.

\subsection{Loss-Landscape of the Solution}
DRL algorithms take the minimization of loss related to Q function $Q(s,a)$ as the main goal of agent training, and these algorithms can be viewed as finding models with loss landscapes that are easier to navigate. A generally accepted view in machine learning is that a flat loss landscape tends to imply better generalization performance \cite{li2018visualizing,chen2022bootstrap,keskar2016large}. An intuitive explanation is that a flat loss landscape means that the model can reach a flat local minimizer, compared to the flat local minimizer, the large sensitivity of the training model at a sharp minimizer negatively impacts the ability of the trained model to generalize on new data. 

Here, the loss-landscape is a plot of the agent loss by perturbing the neural network model parameters $w$ along with the direction of gradients $v$. The loss is calculated on $B$ data points $\{\left( {{s_i},{a_i},{r_i},{s_{i + 1}}} \right)\}_{i=1}^{B}$ with the following formula:
\begin{equation}
\begin{array}{l}
\begin{array}{*{20}{l}}
{{L^\prime }(\theta,\lambda ) = L(\theta + \lambda v) = \frac{1}{B}\sum\limits_{i = 1}^B {{L_Q}\left( {{s_i},{a_i},{r_i},{s_{i + 1}}} \right)}},
\end{array}
 \label{losslandspace_equ}
\end{array}    
\end{equation}
where $\lambda$ is the strength of the perturbation and $L_Q$ is the loss function of the DRL agent.
The loss landscape (\ref{losslandspace_equ}) can indicate whether the trained DRL model could retain a similar loss value when the model parameters are perturbed.

\section{Derivatives of the Digital Twin-Enhanced DRL Solution}

Our proposed digital twin framework can be easily combined with other advanced machine learning techniques to generate derived algorithms. Here we introduce a knowledge distillation algorithm to reduce the complexity of the online bandwidth allocation and an offline DRL method that does not require any online interaction with the real environment.

\subsection{Knowledge Distillation}
We show the proposed digital twin-enhanced DRL solution for the network slicing problem. Although digital twins can help DRL agents better deal with bandwidth allocation problems, the high time complexity of deep neural networks used by DRL agents is still a problem that needs to be alleviated for near real-time network slicing tasks. 
To alleviate the complexity of online policy inference for network slicing, we propose a knowledge distillation scheme.

Knowledge distillation \cite{hinton2015distilling} imparts the capabilities of a teacher model (the original large neural network ) to a smaller student model (the lower complexity neural network) by capturing and ``distilling" the knowledge in the complex neural network model. This student model, allowing for faster inference and lower memory requirements, is easier to apply to real-time demanding tasks without significant loss in performance.

\begin{figure}[t]	\centerline{\includegraphics[width=0.5\textwidth]{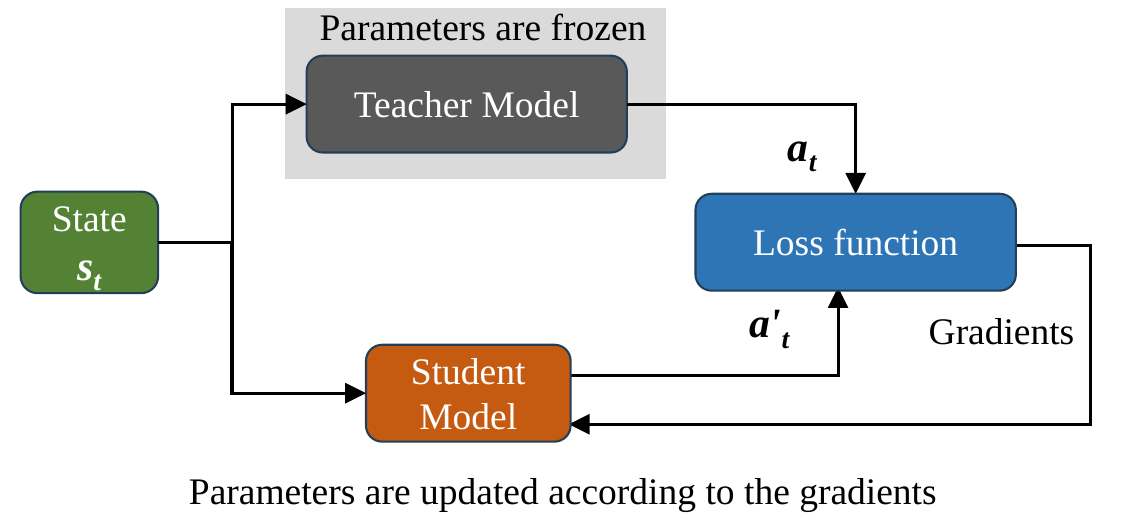}}
	\caption{Knowledge distillation.}
	\label{DistillFramework}
\end{figure}

In our scheme, for the trained DRL model, the data for knowledge distillation can be obtained by sampling from the experience pool. First, the DRL model is regarded as a teacher model $F_T(\cdot,\theta_T)$, and then the historical state $s_t$ is provided as input to the teacher model, and the prediction $a_t$ of the model is used as a pseudo-label. This results in a labeled dataset consisting of states and predicted actions. Further, a loss function can be constructed to train the student model, and knowledge distillation can be realized. A schematic of this scheme is shown in Fig. \ref{DistillFramework}. 

There are 3 steps in the knowledge distillation scheme presented in Fig. \ref{DistillFramework}. First, use the proposed digital twin-enhanced DRL method to train a teacher model. Then, use the data acquisition module to collect some state data $\{s_i\}_{i=1}^{N_d}$, and provide $s_i$ to the trained teacher model $F_T$ to obtain a predicted bandwidth allocation action $a_i=F_T(s_i)$ as a pseudo-label of $s_i$, and then obtain a labeled dataset ${\cal D}_d=\{(s_i,a_i)\}_{i=1}^{N_d}$. Finally, use the gradient descent algorithm to train the student model $F_s(\cdot,\theta_s)$ on dataset ${\cal D}_d$ using (\ref{DistillLoss}). 
\begin{equation}
\mathop {\min }\limits_{{\theta _s}} {L_D}({\theta _s}) = \frac{1}{{{N_d}}}\sum\nolimits_{i = 1}^{{N_d}} {\sum\nolimits_j^{\left| A \right|} {1(j = {a_i})} \log ({p_{ij}})},  
\label{DistillLoss}
\end{equation}
where $1(\cdot)$ is the sign function and $p_{ij}$ is the softmax output of the student model. As the training progresses, the prediction results of the student model will gradually approach the output of the teacher model, i.e., the bandwidth allocation policy of the student model gradually approximates the bandwidth allocation policy of the teacher model. Since the student model is a smaller neural network than the teacher model, it can reduce the complexity of online inference while having the same performance as the teacher model.

\subsection{Offline Solutions}
We claim that the proposed framework for digital twin-assisted intelligent network slicing has a certain universality. Specifically, we generalize this framework to offline reinforcement learning and form a digital twin-enhanced offline reinforcement learning network slicing algorithm. 

In the offline DRL setting, we have a fixed dataset collected by bandwidth allocation policy $\pi_1$ (e.g., a random policy) which is then used to learn new and improved policy $\pi_{off}$ without further interaction with the real environment. We want to use datasets from many past experiences and go beyond naive imitation learning. More formally, in the offline DRL setting, we obtain a static transitions dataset ${\cal D}=\{\left( {{s_i},{a_i},{r_i},{s_{i + 1}}} \right)\}_{i=1}^{M}$ with actions from policy $\pi_1$. In offline DRL, the goal remains the same as in the online case: find a policy that maximizes the expected reward. Therefore, such algorithms often use the static transitions dataset directly to update the parameters of the agent by optimizing the objective function (e.g. the loss function of the DQN algorithm).

Compared with the digital twin-based scheme (Algorithm 1), offline DRL only uses historical data, and it cannot improve exploration because it cannot know whether the data obtained by exploration is valid, whether there is high-quality reward feedback, etc. However, in our proposed framework, during the training of Agent 2, the Agent interacts with the established virtual environment to realize the exploration of strategies, which brings benefits to the performance improvement of the agent. Our proposed Algorithm 1 creates a virtual exploration environment that allows agents to interact and explore, and it allows offline trained DRL agents to interact with online learning DRL agents and also update historical data to calibrate the virtual environment. Here we apply the advantages of digital twins to offline DRL.
Specifically, we propose to replace \emph{Train the DRL Agent 1} in Algorithm 1 with Algorithm 2.

\begin{algorithm}
\caption{Offline Training of the DRL Agent 1}
\label{alg:attack}
    \begin{algorithmic}[1]
    \STATE \textbf{Require:} Batch size $B$, number of training iterations $T$, initialized the neural network model with parameters $\theta_1^{I,1}$, a dataset ${\cal D}=\{(s_i,a_i,r_i,s_{i+1})\}_{i=1}^{M}$.
    \STATE \textbf{For $t=1,\cdots,T$:} \\
        \STATE \quad  Randomly select a batch of data $D_{i = 1}^B \subseteq {{\cal{D}}}$, and
        use the following loss function (\ref{offline_SGDUpdate}) to update the parameters $\theta_1^{I,t}$ related to the Q function and keep the parameter updates of other parts consistent with DRL algorithm $\phi$.
        \begin{equation}
        L'=\frac{1}{B}\sum\limits_{i = 1}^B {{L_Q}\left( {{s_i},{a_i},{r_i},{s_{i + 1}}} \right)}  + \frac{\upsilon }{2}{\left\| {\theta _1^{I,t}  - \theta _2^{I,\Phi}} \right\|^2}.
        \label{offline_SGDUpdate}
        \end{equation}
    \STATE \textbf{End For}
\end{algorithmic}
\end{algorithm}

In Algorithm 2, $L_Q$ is the loss function of the DRL agent.
$\frac{\upsilon }{2}{\left\| {\theta _1^{I,t}  - \theta _2^{I,\Phi}} \right\|^2}$ is a regularization penalty term, and $\upsilon$ is a hyperparameter that controls the degree of penalty. Using this regularization, we penalize our learned value function to make its estimates more conservative.

\section{Simulation Results}

In this section, we investigate the performance of the proposed digital twin-enhanced deep reinforcement learning solutions for resource management in network slicing. First, we compare the proposed approach with the traditional DRL solutions (DQN, DDQN, and Dueling GAN DDQN) and the random solution. Then, we compare the loss landscape of the model trained by the digital twin-enhanced DRL algorithms and the model trained by the traditional DRL algorithms. Then, we show the performance of the knowledge distillation method. Furthermore, we also show the performance of generalizing the proposed framework to offline deep reinforcement learning methods augmented by digital twins.

\subsection{Simulation Setup}

We verify the performance of the proposed digital twin-assisted slicing strategy optimization in a RAN scenario with three types of services (VoLTE, video, and URLLC). Suppose you have the above three corresponding slices in a serving base station. There are 100 slice service rental users randomly distributed within a 40-meter radius around the base station. According to 3GPP TR 36.814, the standard traffic generated by these users is shown in Table \ref{tab:SimulationParameters}. The total bandwidth is 20 MHz, and the bandwidth allocation resolution is 1 MHz or 200 KHz. The neural network of the DQN, DDQN, and the generator of the Dueling GAN DDQN model has 4 hidden layers, each with 256, 256, 128, and 128 neurons, respectively (this neural network structure refers to the open source code in \cite{hua2019gan}). In the considered slice resource management problem, the action space dimension is 4851, that is, the DQN, DDQN, and the generator of the Dueling GAN DDQN model output layer has 4581 neurons. The discriminator of the Dueling GAN DDQN model has 2 hidden layers while each of them has 256 neurons. The optimizer used to train the DRL agent is the Adam optimizer \cite{kingma2014adam} with an initial learning rate of 1e-3. In the simulation results, DQN-DT, DDQN-DT, and Dueling GAN DDQN-DT refer to the digital twin enhanced DRL algorithm obtained by setting $\phi$ in Algorithm 1 to DQN, DDQN, and Dueling GAN DDQN.

\begin{table}
\centering
\begin{minipage}{0.99\linewidth}
\newcommand{\tabincell}[2]{\begin{tabular}{@{}#1@{}}#2\end{tabular}}
\caption{
Simulation Parameters.}
\resizebox{1.0\columnwidth}{!}{
  \centering
  \begin{tabular}{lccccccccccccc}
\toprule
 Parameter  & Value\\
 \midrule
 Bandwidth for all slices & 20MHz\\
 Scheduler & Round robin \\
 Channel model  & Rayleigh fading \\
Distribution of arriving packets of VoLTE & Uniform \\
Distribution of arriving packets of Video & Truncated exponential \\
Distribution of arriving packets of URLLC & Exponential \\
Rate SLA of VoLTE/Video/URLLC & 51Kbps/100Mbps/10Mbps\\
Latency SLA of VoLTE/Video/URLLC & 10ms/10ms/1ms\\
\bottomrule
  \end{tabular}
  \label{tab:SimulationParameters}
}
\end{minipage}\hfill
\end{table}

\subsection{Experimental Results}

\begin{figure*}
\centering
    \subfigure[Reward comparison with $\phi$ is DQN]{
		\begin{minipage}[t]{0.31\linewidth}
			\centering
			\includegraphics[width=1.0\linewidth]{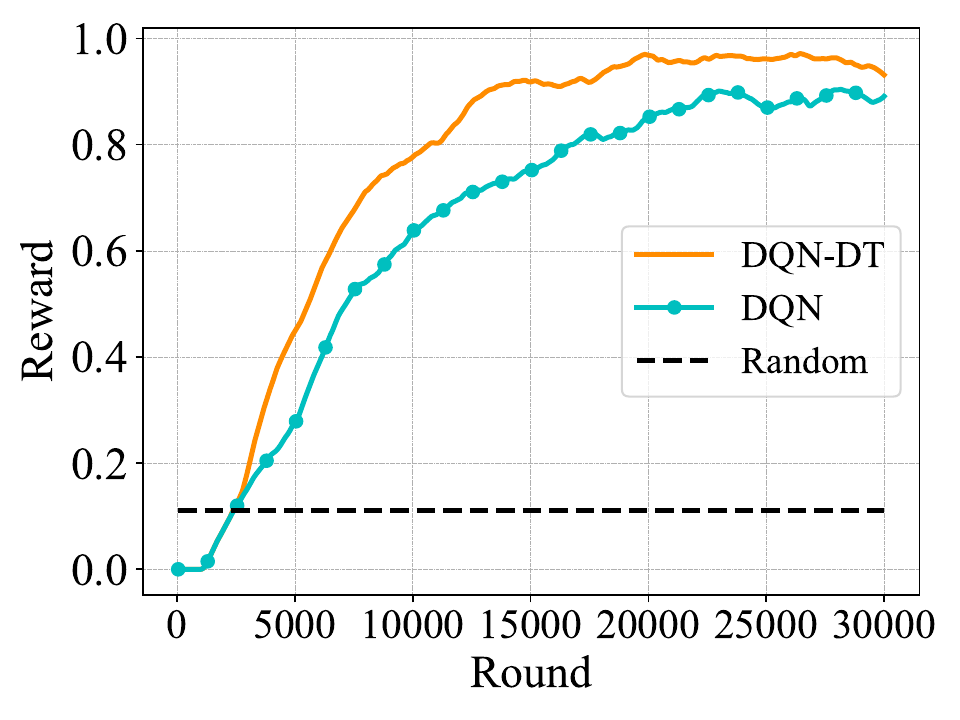}
		\end{minipage}
	}
    \subfigure[Reward comparison with $\phi$ is DDQN]{
		\begin{minipage}[t]{0.31\linewidth}
			\centering
			\includegraphics[width=1.0\linewidth]{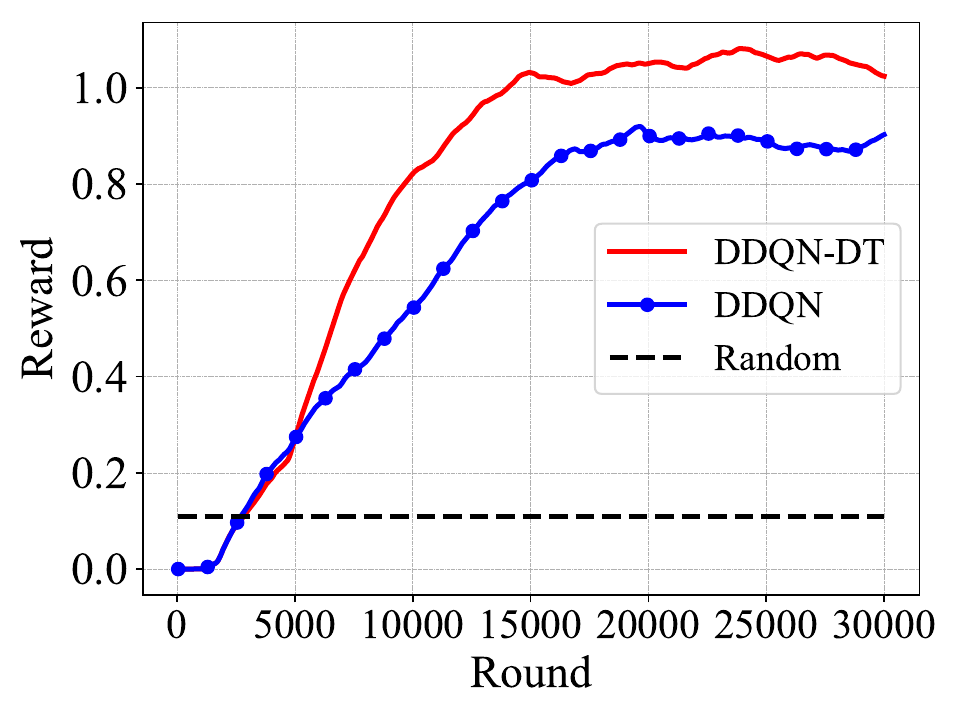}
		\end{minipage}
	}
    \subfigure[Reward comparison with $\phi$ is Dueling GAN DDQN]{
		\begin{minipage}[t]{0.31\linewidth}
			\centering
			\includegraphics[width=1.0\linewidth]{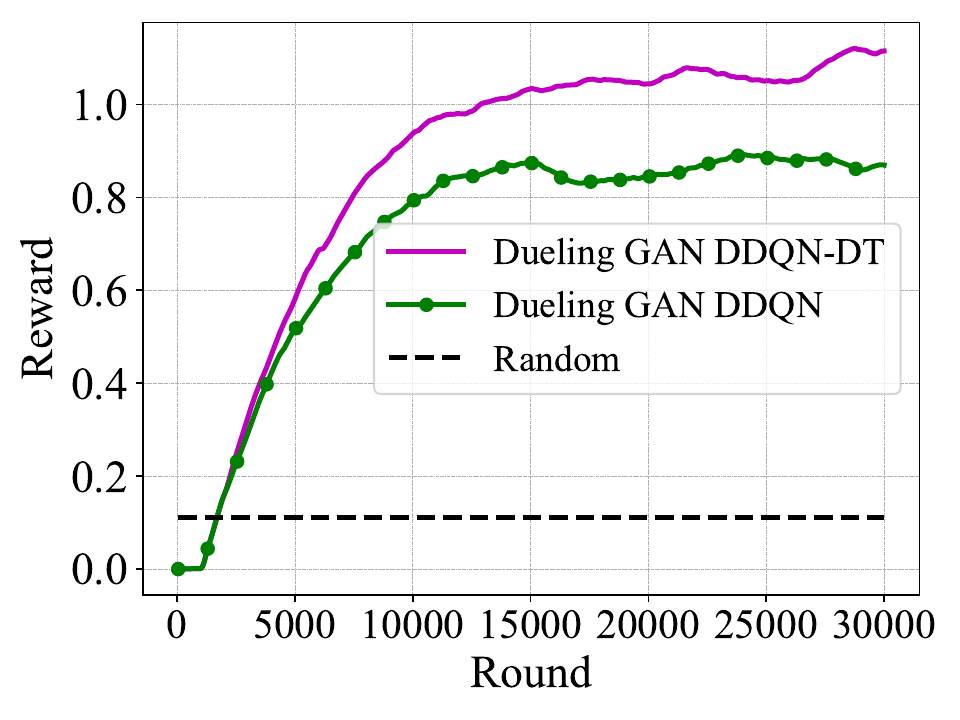}
		\end{minipage}
	}
    \caption{
   Convergence behaviors of the rewards of our digital twin-enhanced DRL solutions and traditional solutions.
    }
    \label{ConvergenceRewards}
\end{figure*}

\begin{figure*}
\centering
    \subfigure[Utility comparison with $\phi$ is DQN]{
		\begin{minipage}[t]{0.31\linewidth}
			\centering
			\includegraphics[width=1.0\linewidth]{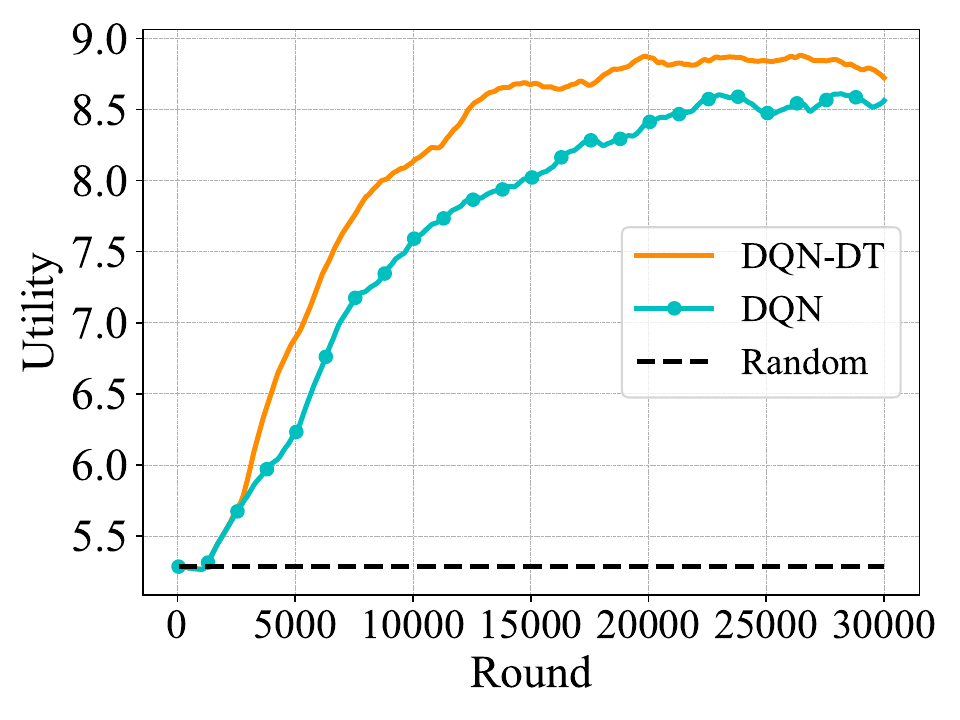}
		\end{minipage}
	}
    \subfigure[Utility comparison with $\phi$ is DDQN]{
		\begin{minipage}[t]{0.31\linewidth}
			\centering
			\includegraphics[width=1.0\linewidth]{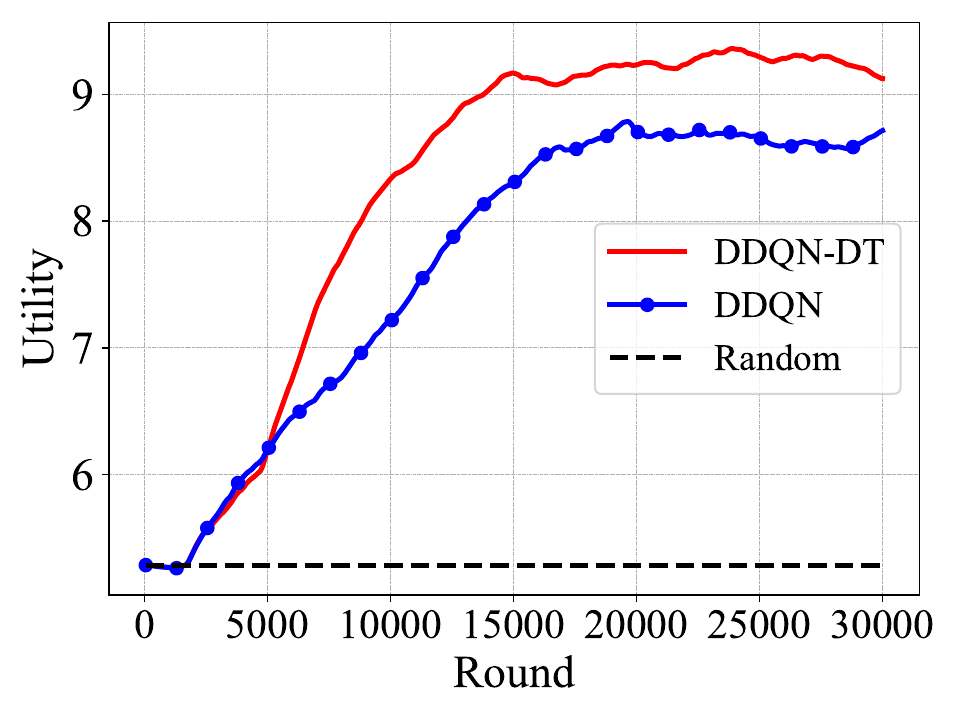}
		\end{minipage}
	}
    \subfigure[Utility comparison with $\phi$ is Dueling GAN DDQN]{
		\begin{minipage}[t]{0.31\linewidth}
			\centering
			\includegraphics[width=1.0\linewidth]{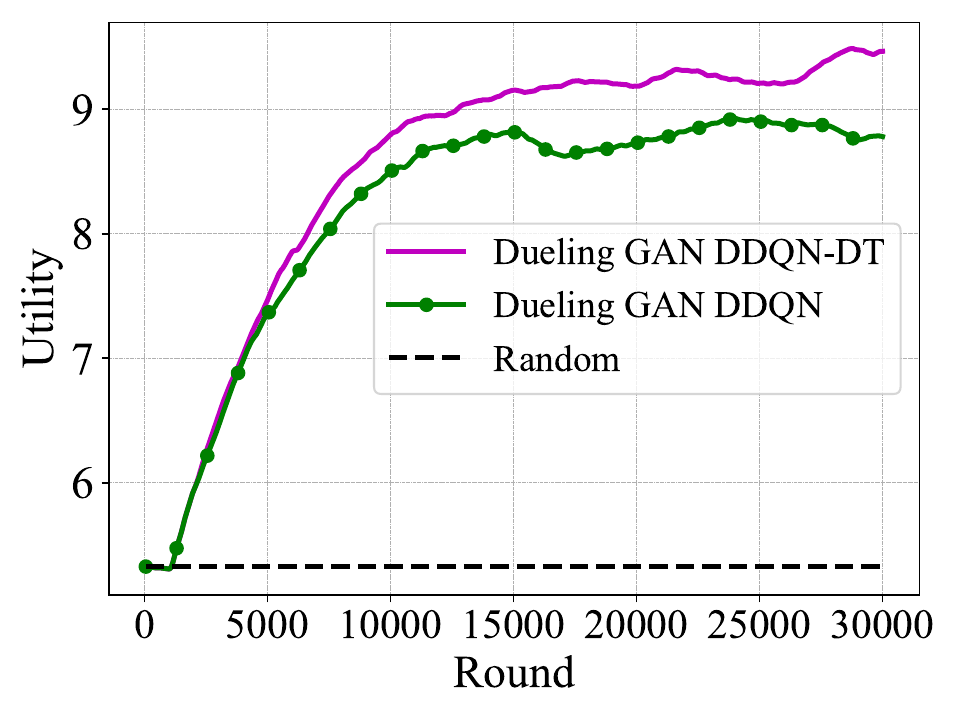}
		\end{minipage}
	}
    \caption{
   Convergence behaviors of the utilities of our digital twin-enhanced DRL solutions and traditional solutions.
    }
    \label{ConvergenceUtility}
\end{figure*}

Here, we compare the convergence curves of digital twin-enhanced DRL algorithms (Algorithm 2) with those of traditional DRL algorithms. In the experiment, we set the agent to interact with the actual network slice environment 200 times as a round, and we show the mean value of the reward (see the equation (\ref{reward_equ})) and the mean value of the utility (see the equation (\ref{utility})) obtained by the agent in each 
round in Fig. \ref{ConvergenceRewards} and Fig. \ref{ConvergenceUtility}. 

First of all, it is clear from Fig. \ref{ConvergenceRewards} and Fig. \ref{ConvergenceUtility} that DRL algorithms enhanced by digital twins have achieved significant gains compared with traditional DRL solutions without digital twins. 
In the experiment, the random policy has the worst effect, because the random policy cannot gain experience from historical data to improve performance, while the DRL algorithm can grasp the law of state transition after interacting with the environment and iteratively optimize better policies. Secondly, it can be found that digital twins speed up the learning efficiency of DRL models, that is, in order to obtain the same reward or the same utility of the two types of algorithms, the digital twin-enhanced DRL algorithms spend fewer training rounds than the traditional DRL algorithms. For example, from Fig. \ref{ConvergenceRewards} (b) we can see that in order for DDQN to obtain a reward value of 0.8, the agent needs 1500 rounds with the environment, whereas our proposed DDQN-DT only needs 1000 rounds.  The above results suggest that our proposed digital twin scheme reduces the number and overhead of real environment interactions required for network slicing to achieve good results. 

\subsection{Loss-Landscape } 
\begin{figure*}
\centering
    \subfigure[Loss landscape comparison with $\phi$ is DQN]{
		\begin{minipage}[t]{0.31\linewidth}
			\centering
			\includegraphics[width=1.0\linewidth]{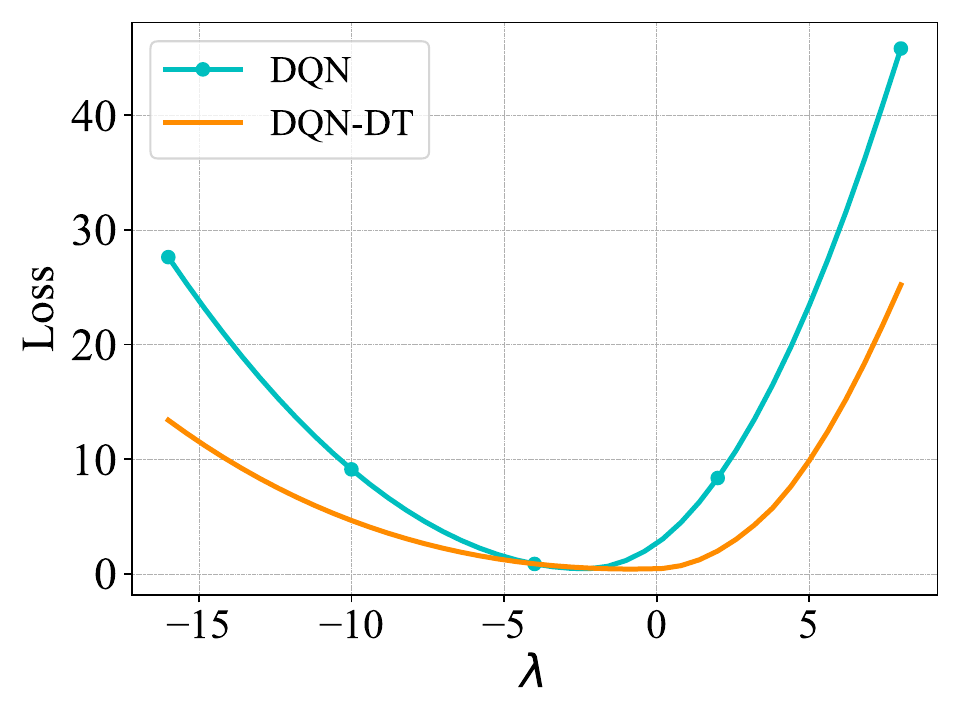}
		\end{minipage}
	}
    \subfigure[Loss landscape comparison with $\phi$ is DDQN]{
		\begin{minipage}[t]{0.31\linewidth}
			\centering
			\includegraphics[width=1.0\linewidth]{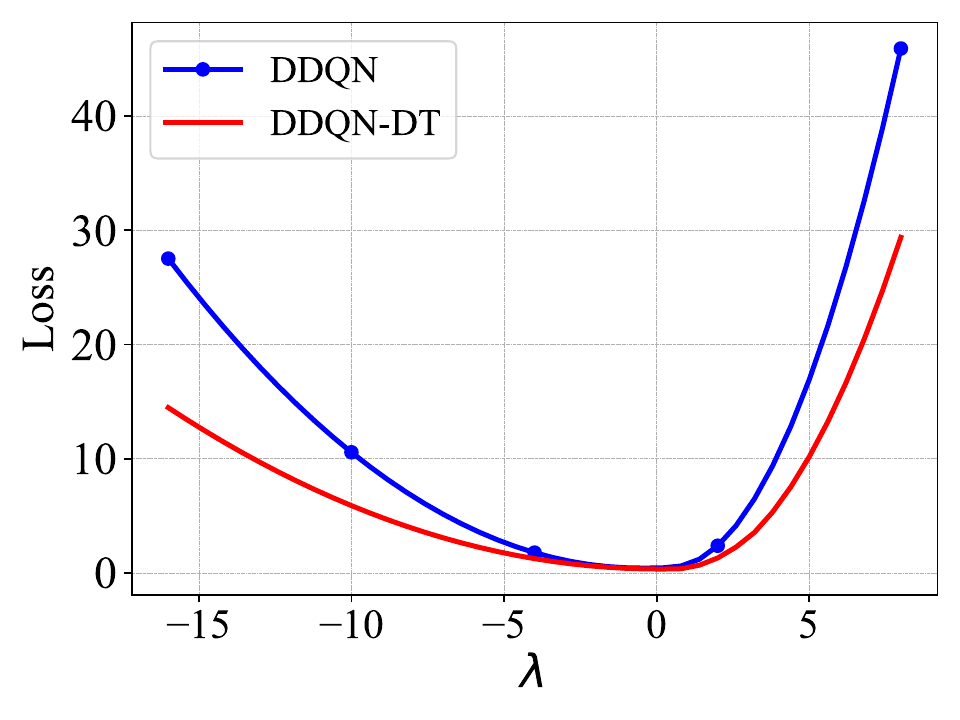}
		\end{minipage}
	}
    \subfigure[Loss landscape comparison with $\phi$ is Dueling GAN DDQN]{
		\begin{minipage}[t]{0.31\linewidth}
			\centering
			\includegraphics[width=1.0\linewidth]{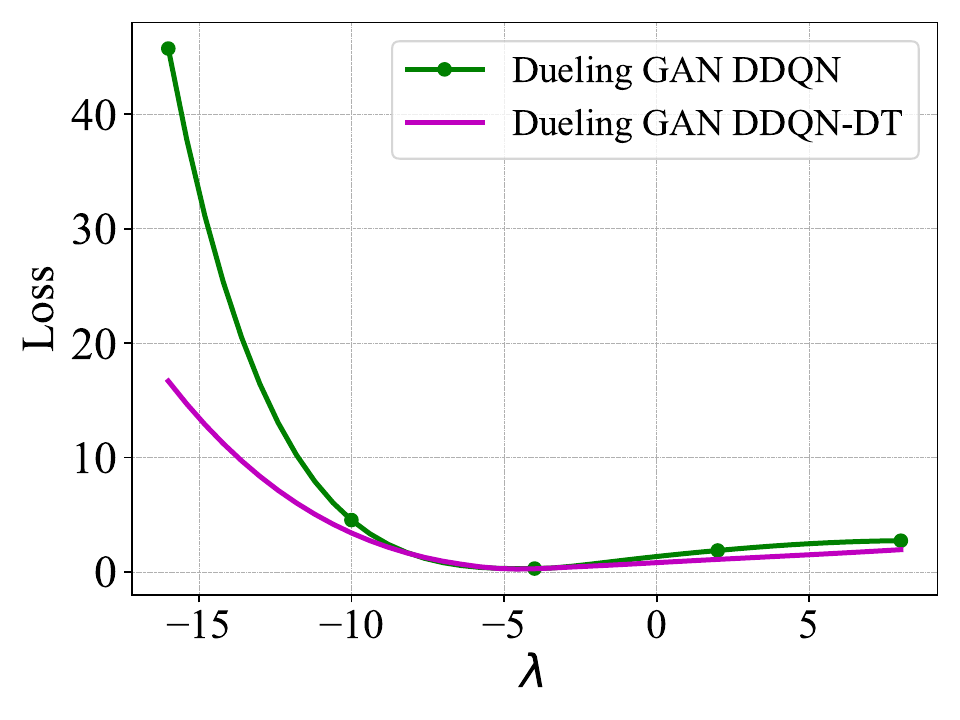}
		\end{minipage}
	}
    \caption{
   Loss landscape perturbed based on the gradient directions of the trained model DRL models.
    }
    \label{losslandspace}
\end{figure*}

In this subsection, first, we compute the gradient directions of the DRL models trained with or without digital twin, and then perturb their parameters along with that direction and measure the loss to plot the loss landscape in Fig. \ref{losslandspace}. The horizontal axis of these plots are perturbations, i.e., $\lambda$ in  (\ref{losslandspace_equ}). The vertical axis is the test loss. 

From Fig. \ref{losslandspace}, we can see that the loss landscapes of DRL algorithms enhanced by digital twins are flatter than those of traditional DRL algorithms. This implies that the weak sensitivity of a model trained by a digital twin-enhanced DRL algorithm at a flat minimum has a smaller generalization error for the training model on new data than at a sharper minimum. This corresponds to the experimental results in the previous section, that is, digital twins improve the generalization ability of DRL algorithms. This also shows that loss landscape is a good indicator for characterizing the generalization performance of network slice optimizer based on DRL.

\begin{figure*}
\centering
    \subfigure[Rewards of DQN and DQN-DT]{
		\begin{minipage}[t]{0.33\linewidth}
			\centering
			\includegraphics[width=1.0\linewidth]{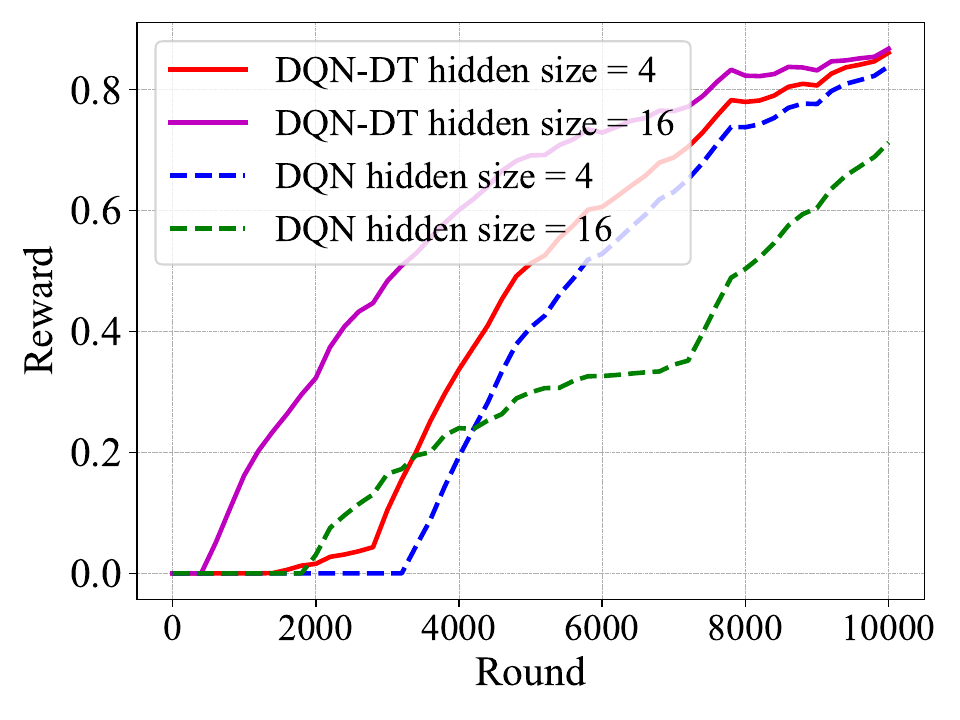}
		\end{minipage}
	}
    \subfigure[Utilities of DQN and DQN-DT]{
		\begin{minipage}[t]{0.33\linewidth}
			\centering
			\includegraphics[width=1.0\linewidth]{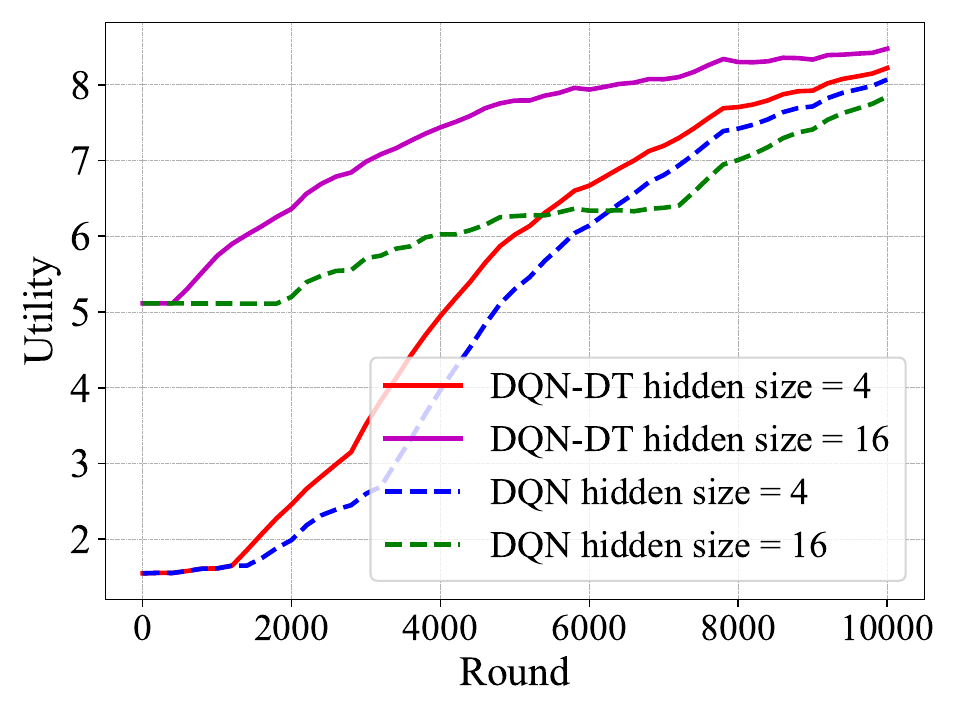}
		\end{minipage}
	}
    \caption{
   Comparison of rewards and utilities with and without the proposed digital twin. DQN-DT means the teacher model is trained by DQN-DT scheme, and DQN means the teacher model is trained by DQN scheme.
    }
    \label{distill_performance}
\end{figure*}

\begin{figure*}
\centering
    \subfigure[Rewards of offline DRL solutions and offline DRL-DT solutions ]{
		\begin{minipage}[t]{0.43\linewidth}
			\centering
			\includegraphics[width=1.0\linewidth]{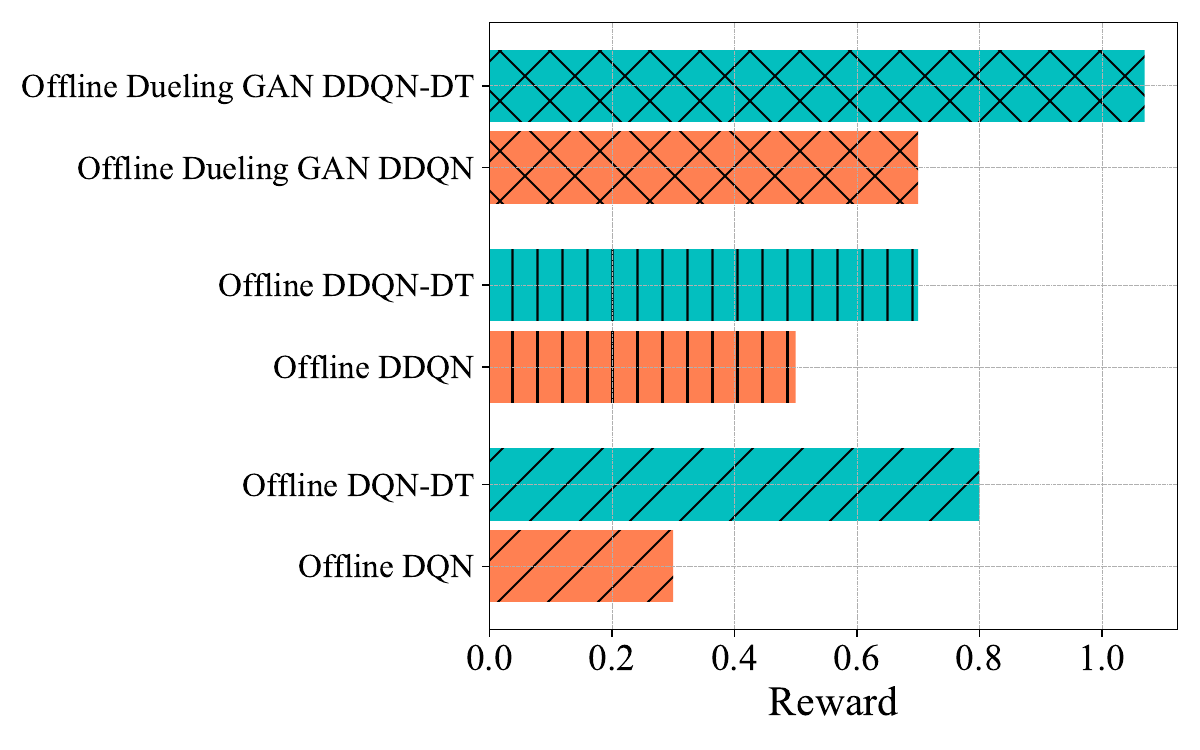}
		\end{minipage}
	}
    \subfigure[Utilities of offline DRL solutions and offline DRL-DT solutions]{
		\begin{minipage}[t]{0.43\linewidth}
			\centering
			\includegraphics[width=1.0\linewidth]{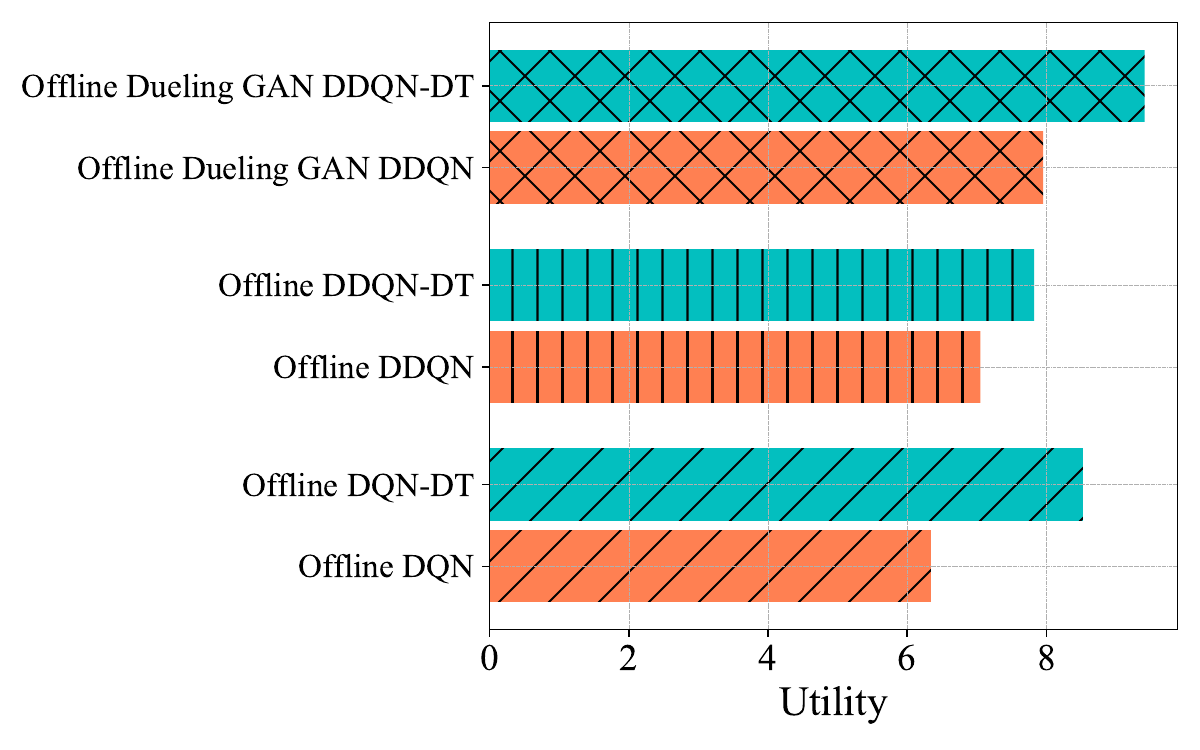}
		\end{minipage}
	}
    \caption{
   Comparison of rewards and utilities of offline DRL solutions with and without the proposed digital twin. 
    }
    \label{Offline_performance}
\end{figure*}

\subsection{Knowledge Distillation}
In this section, we demonstrate the effectiveness of the proposed knowledge distillation method. Specifically, this experiment demonstrates the performance network slicing when the training round is limited and the number of neurons in the student model is much smaller than that in the teacher model (DQN model and DQN-DT model). The purpose of knowledge distillation is to reduce the complexity of network slicing algorithms. To enable the distilled module to be quickly used for network slicing, we set the training budget to 10000 rounds and set the number of hidden layers in the student model to 2, with the number of each hidden layer neurons being 4 or 16. The experimental results are shown in Fig. \ref{distill_performance}.

Fig. \ref{distill_performance} (a) shows the rewards obtained by DQN and DQN-DT when using knowledge distillation, while Fig. \ref{distill_performance} (b) shows the corresponding utility. From these results, it can be seen that under a fixed training budget, even if the hidden layer size of the student model is much smaller than that of the teacher model, the student model can still achieve competitive results through knowledge distillation. In addition, it can be seen that the knowledge distillation effect of the DQN model enhanced by digital twins in the teacher model is significantly better than that of the traditional DQN model in the teacher model. This further confirms the ability of digital twins to improve performance.

\subsection{Results of Offline Solutions}
Here, we show the performance of offline DRL solutions and offline digital twin-enhanced DRL solutions in the considered network slicing scenario. In the experiment, the offline DRL algorithm used as the baseline only used the framework of Algorithm 2 to perform training without interacting with the digital twins in Algorithm 1 framework. We set the number of epochs (traversals of the dataset) for training to 200 and tested the trained model after training. The test results are shown in Fig. \ref{Offline_performance}.

From the experimental results in Fig. \ref{Offline_performance}, we can see that the performance of offline DRL algorithms is not as good as that of online learning DRL algorithms (Fig. \ref{ConvergenceRewards}). This is mainly because DRL agents cannot interact with the actual network slicing environment in any way. However, it is evident from the results that the DRL solutions enhanced by digital twins perform better than the corresponding baseline performance. This indicates that the digital twin module is not only useful for traditional online DRL algorithms but can also be extended to offline DRL algorithms.

\section {Conclusions}
In this paper, we investigated resource management in network slicing scenarios. We leveraged deep reinforcement learning (DRL) to enable intelligent resource management in dynamic scenarios with uncertainty. We proposed a digital twin-enhanced DRL method to improve the learning efficiency of the DRL agents. Experimental results demonstrated that the proposed digital twin solution has performance advantages and can reduce the number of interactions between the agent and the real environment. 
We also proposed using loss landscapes to visually analyze the performance of traditional DRL-trained models and digital twin-enhanced DRL-trained models. The experimental results indicated that the model trained by the digital twin-enhanced DRL algorithm has a flatter loss landscape. In addition, the knowledge distillation method proposed in this paper can be used to alleviate the complexity of network slicing models based on deep neural networks. Besides, our proposed offline solution confirms that our proposed digital twin framework can be extended to offline DRL algorithms.

In the future, we will test the proposed digital twin frame under a broader deep reinforcement learning framework to verify its practical value and extend our work to other wireless communication issues. In addition, theoretically analyzing the role played by digital twin modules is also a direction worthy of future research.
Besides, one can notice that the LSTM we designed to model state transitions can be thought of as a state generator. Predicting the next state is like the classic natural language processing task of predicting the next word. One perspective is that we can think of the LSTM in the digital twin as a ``chat robot". The agent communicates with the chat robot and improves its ability to obtain rewards through feedback. A natural idea is that in the future we can try to use large language models (LLMs) to build a digital twin to improve overall performance. In the future, we will explore the construction and efficient training and inference of digital twins based on LLMs.

\ifCLASSOPTIONcaptionsoff
  \newpage
\fi

\bibliographystyle{IEEEtran}
\bibliography{IEEEabrv,DRL}

\end{document}